\documentclass{PoS}

\usepackage{subfig}

\title{Feasibility of VHE gamma ray detection by an array of imaging atmospheric Cherenkov telescopes using the fluorescence technique}

\ShortTitle{VHE gamma ray detection with ACTs and fluorescence}

\author{J.L. Contreras, J. Rosado, F. Arqueros,M. L\'{o}pez, J.A. Barrio and M. Nievas\\
        Departamento de F\'{i}sica At\'{o}mica, Molecular y Nuclear, Facultad de Ciencias F\'{i}sicas,
        Universidad Complutense de Madrid, E-28040 Madrid, Spain\\
        E-mail: \email{marcos@gae.ucm.es}}

\abstract{The last 20 years have seen the development of new techniques in Astroparticle Physics providing access to the highest end of the electromagnetic spectrum. It has been shown that some sources emit photons up to energies close to 100 TeV. Yet the fluxes of these photons are incredibly low and new detection techniques are needed to go higher in energy.

A new technique that would use the new generation of Cherenkov Telescopes, i.e., the Cherenkov Telescope Array (CTA), is proposed to push further the energy frontier. It is based on the detection of the fluorescence radiation emitted in extensive air showers, a successful method used in ultra-high-energy cosmic ray experiments, like the Pierre Auger Observatory. It would complement the standard imaging atmospheric Cherenkov technique with only minor modifications of the hardware currently being developed for the CTA and would not imply significant extra costs during its planned operation.

 }

\FullConference{The 34th International Cosmic Ray Conference,\\
		30 July- 6 August, 2015\\
		The Hague, The Netherlands}

\begin{document}

\section{Introduction}

Cherenkov telescopes as MAGIC, H.E.S.S. or VERITAS point to the suspected sources of very high energy (VHE) photons looking for the Cherenkov radiation of the extensive air shower (EAS). They cover small fields of view (tens of square degrees) over a range of energies that spans from some $10^{10}$ to $10^{14}$ eV \cite{hegra-crab}, \cite{MAGIC-crab}. On the other hand, fluorescence telescopes rely on the detection of the isotropic air fluorescence generated by the charged particles of the EAS. This technique has been successfully used in ultra-high-energy cosmic rays experiments (e.g., Pierre Auger Observatory) with a threshold close to $10^{17}$ eV and therefore well above the VHE domain \cite{Auger-fluo}. In this work we  propose to extend this technique to lower energies using the large light collection areas available in arrays of Cherenkov telescopes (ACTs).    

Cherenkov astronomy is bound to change with the construction of the Cherenkov Telescope Array (CTA), presently in its design and prototyping phase \cite{Introd-CTA}. CTA will be composed of two observatories, one in each hemisphere. Each observatory will consist of a large array of Cherenkov telescopes made up of small, medium, and large size reflector dishes of 4, 12 and 23 meters diameter, respectively. The Southern observatory will be the largest, housing around 100 telescopes and extending over more than 1 km$^2$. Its sensitivity will be a factor of 10 better than existing imaging atmospheric Cherenkov observatories and its energy range will cover from 20 GeV to over 200 TeV. The first telescopes are expected to be installed in 2017.

CTA has been designed as a pure Cherenkov observatory. Our goal is to study the possibility 
to operate CTA and similar ACTs as mixed Cherenkov and fluorescence observatories. We think that ACTs operating in fluorescence mode could reach a similar effective area as in Cherenkov mode,
although with a much higher threshold, of the order of 100 TeV. On the other hand they would be able to detect showers from a very wide angular region, more than one stereoradian. The fluorescence and Cherenkov modes would coexist at no extra observation cost. The combination of both techniques would extend ACT capabilities in the high energy range to be able to reach the 1 PeV region. In the following we detail these estimates.

\section{Toy model}

We have developed a toy model to roughly estimate the effective areas that an ACT, with characteristics similar to those of the the sub-array composed by the Medium Size Telescopes of CTA, could reach in fluorescence mode. The model is defined by the following approximations:

\begin {itemize} 
  \item Only gammas are considered as primary particles. We have not performed estimations for protons or other nuclei.
  \item Interaction of primaries with the atmosphere is simulated by a parameterization of the energy deposited by the shower as a function of atmospheric depth, using a Greisen profile.  
  \item Showers are assumed to have negligible transverse spread. 
  \item Impact points of the showers on ground are distributed on a square grid of 100 m spacing.
  \item Fluorescence intensity is proportional to the energy deposition according to the (pressure, temperature and humidity dependent) fluorescence yield \cite{fluo-yield}. The atmospheric transmission is calculated for each line of the fluorescence spectrum.
  \item Telescopes are simulated as ideal detectors with uniform efficiency, around 35\%, in the relevant wavelength range.
  \item A telescope is assumed to be triggered when it detects more than 200 photo-electrons. 
  \item Only telescopes similar to CTA Medium Size Telescopes (MSTs) are considered. They are described as 12 m diameter reflectors with a field of view (FoV) of 8 degrees diameter.
  \item We consider 24 telescopes spread on an area similar to the one taken in CTA design (a radius of 300 m).
  \item All the telescopes point to the same direction.
  \item A shower is considered as detected when it triggers at least one telescope.
\end{itemize}

\begin{figure}%
\centering
\subfloat[Telescope positions simulated.]{{\includegraphics[width=0.4\linewidth]{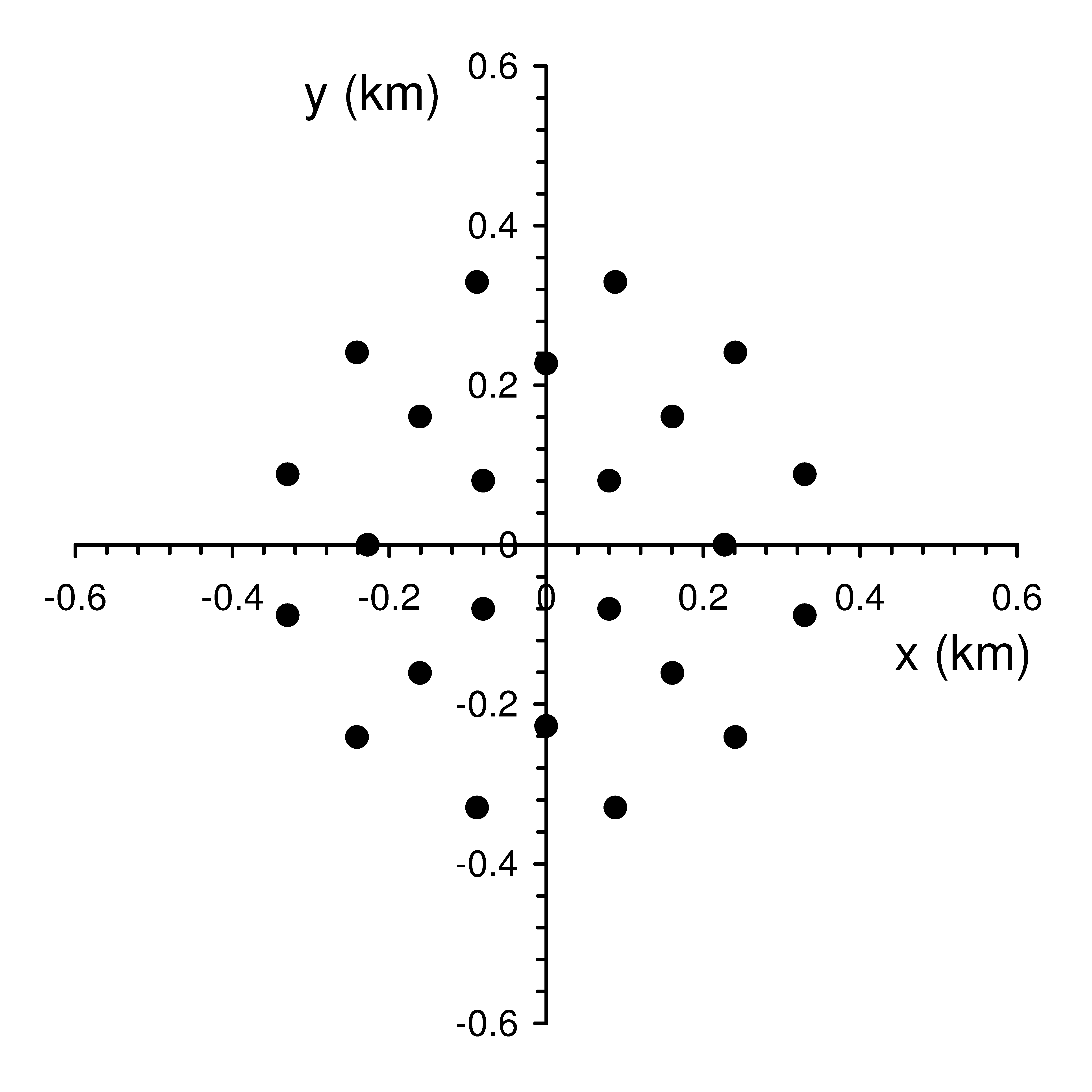}}}%
\qquad
\subfloat[Profile of a 160 TeV shower.]{{\includegraphics[width=0.4\linewidth]{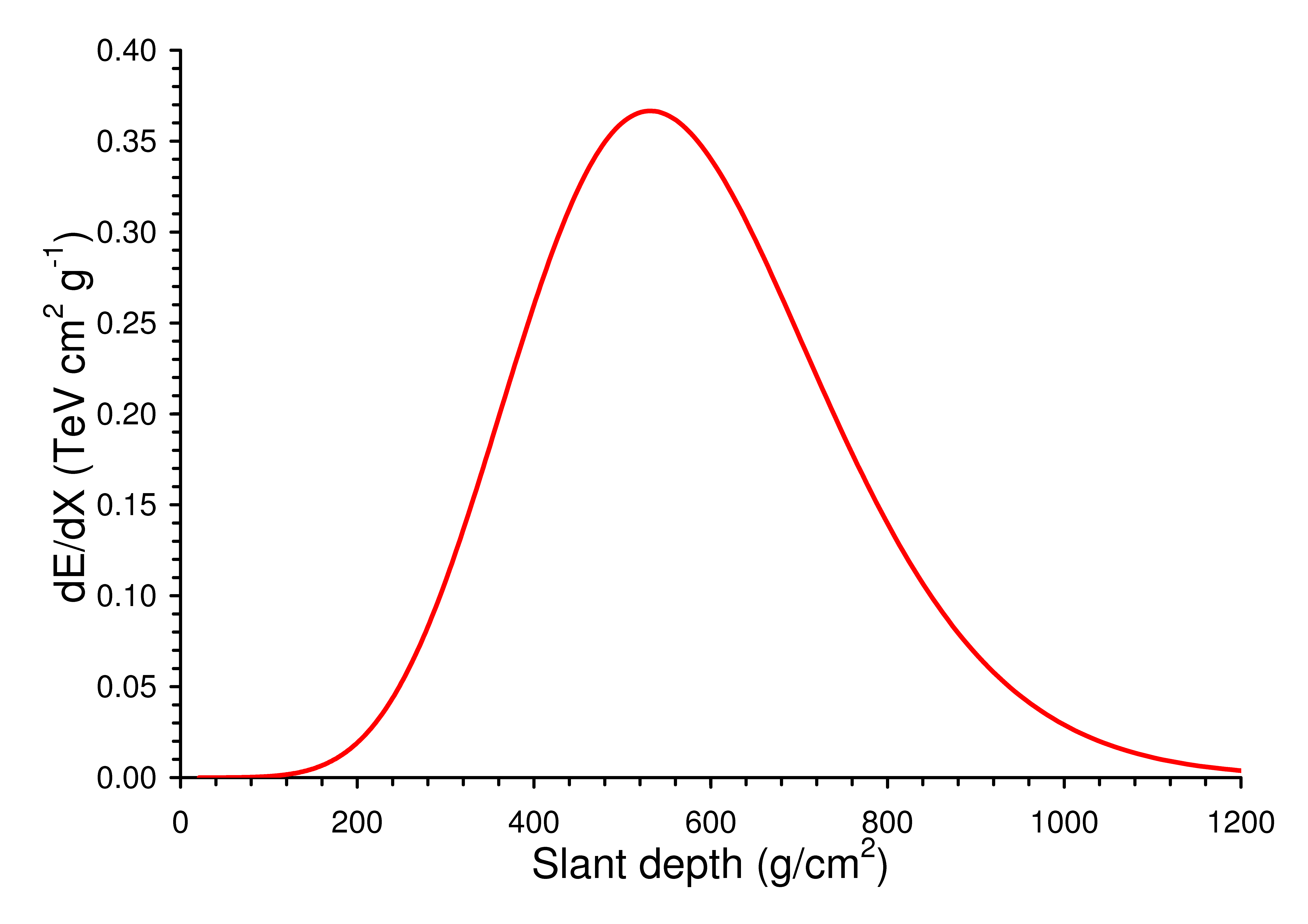}}}%
\caption{Some of the ingredients of the toy model used.}
\label{fig:profile}
\end{figure}

Figure \ref{fig:profile} shows the layout of the telescopes (left hand) and the energy deposition versus atmospheric depth assumed for a shower of 160 TeV (right).

\begin{figure}
\centering
\includegraphics[width=1.0\textwidth]{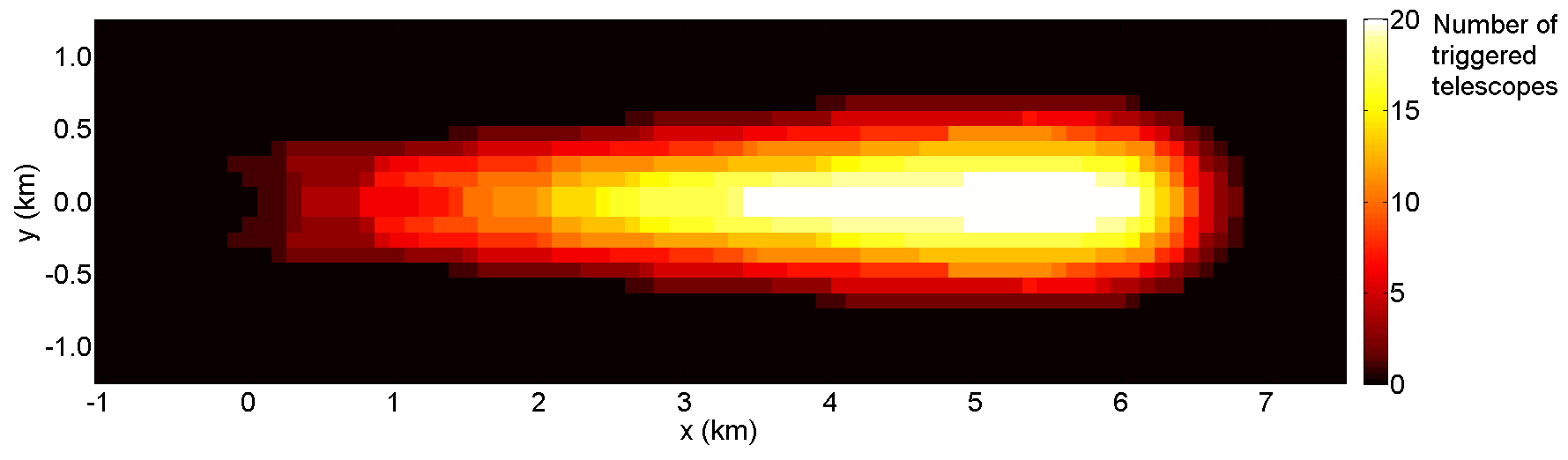}
\caption{Geometrical representation of the calculation of the effective area for 1 PeV showers at 40$^\circ$ zenith angle.
 The array center is located at the origin.  The colour axis represents the number of telescopes triggered by the shower.}
\label{fig:footprint}
\end{figure}

  The effective area can be visualized by means of a geometrical construction. Let us imagine an inverted
truncated cone based at the array and extending upwards. It represents the field of view of the array. Showers 
inside the volume of  this truncated cone can be detected if they leave enough light on the telescopes. After crossing the volume they will
impact on ground. Given an incoming direction for the showers (the direction of the source) the distribution of the
impact points (core positions) for the detected showers will be a sort of ``shadow'' of the FoV. The area of this shadow
approximates the effective area for those conditions. The simulations performed have been based on this image.
This is illustrated for a 1 PeV shower and 40$^\circ$ zenith angle in figure \ref{fig:footprint}.

Usually ACTs operating in Cherenkov mode demand that at least two telescopes are
triggered by a shower. Therefore our last condition seems too loose. Nevertheless, in our simplified approach, this is not relevant, since the number of showers seen by just one telescope is less than 10\%, as can be checked in the example shown in the figure.

\section{Results}

\begin{figure}
\centering
\includegraphics[width=.8\textwidth]{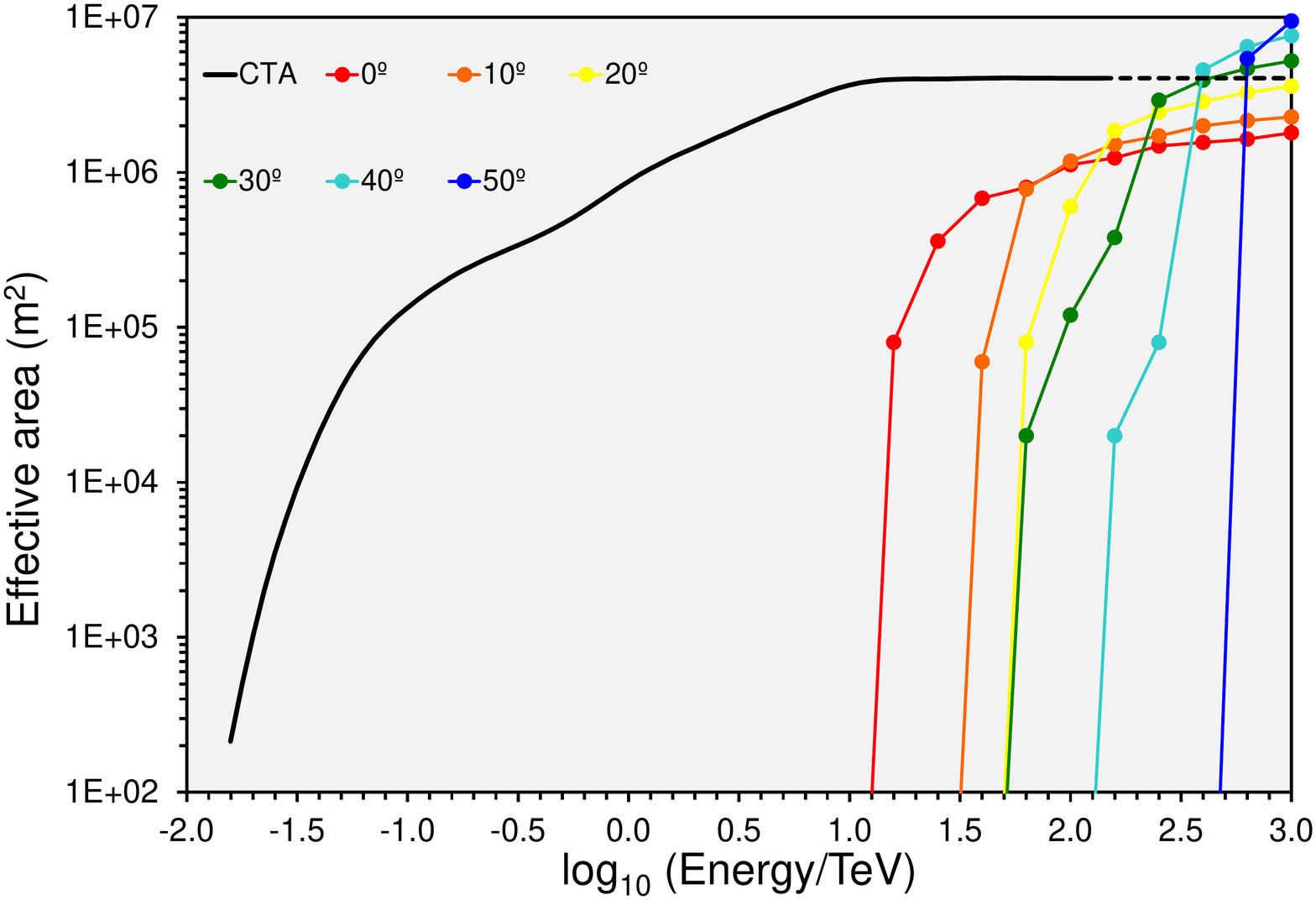}
\caption{Effective area for fluorescence mode predicted by the toy model, as a function of energy for different zenith angles of the source. The effective area of CTA South in Cherenkov mode, taken from reference \cite{cta-performance}, is also shown for comparison.}
\label{fig:EffArea200}
\end{figure}

In order to estimate the effective areas for gamma-rays two different viewing directions of the array were assumed: a first one with the telescopes pointing to the zenith and a second one with an inclination of $45^\circ$. They led to similar results. Figure \ref{fig:EffArea200} summarizes our results for the first case. The colored lines represent the effective areas obtained from our toy model for the detection of gamma rays as a function of their energy, for different zenith angles of incidence of the shower. The continuous line depicts the effective area computed for the Southern observatory of CTA after quality cuts are applied, obtained from \cite{cta-performance}. It has been extrapolated (dashed line)  from the highest energy found in the reference (around 200 TeV) to 1 PeV assuming a constant value, following the tendency from the previous points.

  We can summarize the results in the following numbers:

\begin{itemize}
\item With the trigger condition imposed fluorescence begins to contribute around 50 TeV.
\item The threshold increases with the zenith angle and also the effective area at very high energies. For energies above 400 TeV and angles above 40$^\circ$, the effective areas obtained with the proposed method are larger than 5 km$^2$, exceeding those obtained with the standard Cherenkov mode, marked in the figure, which is around 4 km$^2$ \cite{cta-performance}, the size of the full array.
\end{itemize}

The main trends present in the figure can be readily understood from the discussion in the previous section.
When the direction of the shower forms a small angle with the pointing direction of the telescopes
a large fraction of the shower development, and therefore the light emitted,  is contained within the FoV.
In consequence the energy threshold for detection decreases with the angle between both.
On the other hand, as the angle increases so does the size of the ``shadow'' and thus the effective area.

Not shown in the figure is the aperture. It depends on both the energy of the shower and the pointing angle of the array. A first approximation can be readily computed using the toy model. The result is that if we consider that the telescopes have an effective area of 2-3 km$^2$ to detect 200 TeV gamma rays with incidence angles below 40$^\circ$ (i.e., about 25\% of the visible sky), the aperture would be 3-4 km$^2$ sr. This is more than 30 times the Cherenkov aperture at the same energy. At higher energies, the aperture would be even much larger as both the effective area and the sky coverage for fluorescence keep growing.

Many extra hours of visibility per year and source could be obtained if fluorescence mode were available during regular Cherenkov observation. For example, for the MAGIC observing site, located at Observatorio del Roque de los Muchachos in the island of La Palma, a source passing through the zenith, as the Crab Nebula, can be observed during dark time and at zenithal angles lower than 40$^\circ$ for more than 400 hours per year, assuming that the array is always pointing to the zenith. This extra time would be welcome because the expected flux for the Crab Nebula over 100 TeV is somewhat below 0.5 photons/km$^2$/h, indicating the large effective areas and observation times needed to detect Pevatrons.

Obviously the effective areas estimated depend strongly on the assumption of 200 photo-electrons threshold imposed. Lower cuts, of the order of 50 photo-electrons, are usually imposed in Cherenkov mode (see for example \cite{MAGIC-crab} and references within). Our number takes into account that the light would be spread over a larger section of the camera and that the fluorescence light pulse collected in a pixel would be in general longer than the Cherenkov one. This last factor depends on the relative orientation of the shower axis with respect to the telescope axis. When the shower and telescope axis are anti-parallel the fluorescence light follows the same direction as the shower front compressing the arrival time interval. In our simulation the longest development times of triggered showers in the camera were around 4 microseconds, corresponding to about 80 ns pulse length in a given pixel. They originate in distant showers crossing the field of view at angles nearly perpendicular to the telescope axis. Registering these signals could require to adapt the detection threshold as a function of the relative orientation of the shower, taking more restrictive thresholds for perpendicular showers.

Another important point is that in this simplified approach we have not emulated the full CTA array. The layouts presently considered for the Southern observatory include 4 Large Size Telescopes (LSTs) and 72 Small Size Telescopes (SSTs). If these were included in the model for sure they would reduce the threshold, due to the larger mirror area and reduced FoV of the LSTs (23 m), and extend the effective area at the highest energies, due to the large number of SSTs.

\section{Open points}

The very preliminary results shown only intend to prove the feasibility of the proposal. Many points remain open and have to be studied before the result is established.

First of all, operating CTA in fluorescence mode would need of changes in the hardware of the telescopes. The reason is that, as we already mentioned, the fluorescence light pulse is much longer than the Cherenkov one, especially when the shower forms a right angle with the telescope axis. It must be studied if the camera alternatives proposed for CTA  MSTs (see \cite{flashcam} and \cite{nectarcam}) can be adapted to cope with this conditions. Limitations on the acceptable pulse length and shower development time in the camera would translate in a lower range of acceptable incidence angles; stated differently, in a smaller fluorescence field of view. Design studies for CTA have proposed trigger and readout systems that follow the development of the image of showers in the camera, as the one described in \cite{colibri}. Thought as a solution for following the development of Cherenkov images created by showers at large impact parameters, it could be applicable to our more extreme case with longer development times.

In the second place the threshold of 200 photo-electrons, imposed as simplified trigger condition, has to be cross-checked, translated into a meaningful trigger requirement and applied in simulations that include Cherenkov events, night sky background and detector response.

We also note that the proposed technique would change the philosophy of the array operation, at least for the highest energies. While in Cherenkov mode observations are planned pointing to the targets of interest, if fluorescence mode is enabled high energy photons would be detected from many different sources during the same observation. Only offline would their directions be disentangled and photons classified according to their origin. This would also change the way to unfold the instrumental response, introducing dependencies on new variables. 

Reconstructing the direction of the photons should not be a problem, as a direct extension of the method actually employed in fluorescence telescopes, based on image and timing, should work \cite{Auger-fluo}. Although ACTs will register a shorter shower path length, they will reach higher geometrical accuracy. In regard with the energy reconstruction, the fluorescence technique achieves presently resolutions of the order of 25\% on the energy scale, but to do so it profits from the reconstruction of the whole longitudinal development of the shower. ACTs would only observe a small fraction of the development, from which the energy of the shower would have to be extracted. It is not clear yet how to achieve this. The fact that several telescopes would observe the shower at the same time would certainly help for the direction reconstruction and probably also for the energy one. 

Finally a key point to make this proposal worthy would be the ability of the technique to separate gammas from other cosmic rays. The latter are more than 100 times more numerous than gammas. Even without a separation method the extra aperture would add value to ACTs as cosmic ray observatories (something which is non negligible), but not as a PeV astronomy observatory. In fluorescence detectors, the separation between different primaries is based on the longitudinal profiles of showers, but as mentioned before, only a part of the profile would be registered due to the limited FoV of ACTs. On the other hand the precision on the lateral distribution would be much better due to the 20 times higher angular resolution of the ACT cameras. Gamma shower lateral profiles are more compact and more luminous. If methods are found that can profit from it, a good separation would be achievable. Right now we cannot estimate its rejection power because it needs a detailed Monte Carlo study. 

\subsection{Added value of the proposal}

Besides the increased aperture and its impact on the possibility of building PeV gamma-ray astronomy, which are the key goals of this proposal, other points add value to it. We briefly enumerate them:

\begin{enumerate}
 \item It could extend the ACT cosmic ray programs.
 \item In some conditions fluorescence observations could be continued through horizontal observations at cloudy nights.
 \item A proper implementation of fluorescence radiation and Cherenkov light scattering in the simulation of ACTs would allow a better evaluation of the fluorescence contamination in Cherenkov events, allowing to reduce or put limits to the associated systematic uncertainty in the reconstruction of primary gamma rays.
\end{enumerate}

\section{Conclusion}

 We have studied the feasibility of observing fluorescence radiation from VHE gamma-ray showers using an array of Cherenkov telescopes with characteristics similar to those proposed for a sub-array of CTA. Using a simple toy model and trigger conditions we find that the array would be able to detect gamma-ray showers from $\sim 100$ TeV with effective areas above a few square kilometers, covering about 1 sr of the sky at any given moment. At the same time the system could open the door to a new cosmic ray program using ACTs.
 
 The proposal would imply modifying the electronics and operation mode presently used in ACTs and it is not clear if a trigger and data acquisition scheme could be easily implemented to serve it. It is not easy either to foresee if efficient gamma/hadron separation algorithms would be feasible using the acquired information. Nevertheless, we think it is worthwhile to perform a full Monte Carlo and technical study of this possibility, which could be considered in a future CTA upgrade.  

\section*{Acknowledgements}

This work was supported by the Spanish MINECO under contract FPA2010-22056-C06-06, FPA2012-39489-C04-02 and FPA2013-48381-C6-2-P.

\end{document}